\documentclass[sigconf]{acmart}

\AtBeginDocument{%
  \providecommand\BibTeX{{%
    \normalfont B\kern-0.5em{\scshape i\kern-0.25em b}\kern-0.8em\TeX}}}

\begin{document}

\title{CONTAIN: Privacy-oriented Contact Tracing Protocols for Epidemics}

\author{Arvin Hekmati}
\affiliation{%
  \institution{University of Southern California}
  \city{Los Angeles}
  \state{CA}
  \postcode{90089}
}
\email{hekmati@usc.edu}

\author{Gowri Ramachandran}
\affiliation{%
  \institution{University of Southern California}
  \city{Los Angeles}
  \state{CA}
  \postcode{90089}
}
\email{gsramach@usc.edu}

\author{Bhaskar Krishnamachari}
\affiliation{%
  \institution{University of Southern California}
  \city{Los Angeles}
  \state{CA}
  \postcode{90089}
}
\email{bkrishna@usc.edu}

\begin{abstract}
Pandemic and epidemic diseases such as CoVID-19, SARS-CoV2, and Ebola have spread to multiple countries and infected thousands of people. Such diseases spread mainly through person-to-person contacts. Health care authorities recommend contact tracing procedures to prevent the spread to a vast population. Although several mobile applications have been developed to trace contacts, they typically require collection of privacy-intrusive information such as GPS locations, and the logging of privacy-sensitive data on a third party server, or require additional infrastructure such as WiFi APs with known locations. In this paper, we introduce CONTAIN, a privacy-oriented mobile contact tracing application that does not rely on GPS or any other form of infrastructure-based location sensing, nor the continuous logging of any other personally identifiable information on a server. The goal of CONTAIN is to allow users to determine with complete privacy if they have been within a short distance, specifically, Bluetooth wireless range, of someone that is infected, and potentially also when. We identify and prove the privacy guarantees provided by our approach. Our simulation study utilizing an empirical trace dataset (Asturies) involving 100 mobile devices and around 60000 records shows that users can maximize their possibility of identifying if they were near an infected user by turning on the app during active times.
\end{abstract}




\maketitle

\section{Introduction}
Viruses such as CoVID-19 and SARS-CoV2 have infected hundreds of people throughout the world, resulting in multiple fatalities. The World Health Organization has issued guidelines to limit virus outbreak by tracking people who may have come in contact with an infected individual~\cite{world2014contact}. Such measures allow government and health care officials to drastically reduce the spread of the virus to a large community. The infected individual is required to share all his or her travel details with the health care authorities to reliably track and quarantine people who may contract the virus due to their physical proximity with the infected individual.

The contact tracking process typically involves gathering privacy sensitive information from the infected individual. Privacy-savvy individuals may not be willing to share all the information, which may hamper the contact tracing process while exposing the vast population to continued spread. A privacy-sensitive contact tracking approach would encourage the people to participate in the contact tracing process confidently, which is the focus of this work.

Several contact tracking applications involving mobile applications, wireless technologies, and GPS have been presented in the literature~\cite{danquah2019use,farrahi2014epidemic,10.1145/2820783.2820880,8422886,prasad2017enact,7763193,DBLP:journals/corr/ReddyKRC15,YONEKI201483}. Such approaches either expect the infected individual to self-report their contact information or rely on external infrastructures such as wireless access points and GPS. On the one hand, the approaches requiring self-reporting are not reliable since the individuals may not honestly report the information to the government authorities. On the other hand, the infrastructure-dependent methods work only when the mobile device is in the proximity of the infrastructure; note that the GPS reception is not reliable in indoor environments. EPIC~\cite{8422886} and ENACT~\cite{prasad2017enact} were developed to trace contacts using a mobile application in a privacy-preserving manner. While they avoid the use of GPS, these frameworks rely on a wireless access point. Mobile phones report the access point identifier and the timestamp to a server, which is then matched against the infected user's information to detect contact proximity.


In this work, we present CONTAIN (an acronym coined from ``CONtact TrAcINg"), which is a mobile application to enable privacy-sensitive contact tracing for epidemics. We show how to leverage the users' mobile devices sending anonymous encrypted or random messages to each other via Bluetooth to allow users to determine with a 100\% privacy if they have been in range of an infected user in the past, and when. There is no collection of privacy sensitive GPS, no  reliance on external infrastructure for location tracking, no continuous logging of personally identifiable information from users. Specifically we introduce two privacy-friendly contact tracing protocols for CONTAIN; The first protocol uses symmetric key encryption, while the second protocol requires the secure generation of sufficiently large random numbers. In a way, they are duals of each other: the first protocol nodes upload the messages they have heard, while in the second protocol, the nodes upload the messages they have sent. Both protocols provide privacy for users that are not infected as such users never need to reveal their identity or contacts. Both protocols offer users with opt-in measures to get verified and declare if they are infected, while still maintaining some measure of privacy for the general public as they do not need to reveal their true identities or those of their contacts or the locations they have visited to everyone. And both protocols allow users to privately detect if they have had contacts with others that are infected.


\section{Related Work}
\label{sec:rw}
During the EBOLA outbreak in 2014, the World Health Organization (WHO) explained the importance of contact tracing and laid down protocols for tracing contacts~\cite{world2014contact}. But, WHO did not employ any mobile application, and they have only provided guidelines for the health care workers to improve the efficiency of the contact tracing process. A number of mobile phone based contact tracing applications have been presented in the literature~\cite{danquah2019use,farrahi2014epidemic,10.1145/2820783.2820880,8422886,prasad2017enact,7763193,DBLP:journals/corr/ReddyKRC15,YONEKI201483}. 

Danquah~\textit{et al.}~\cite{danquah2019use} presents an electronic system for contact tracing to address the shortcomings of paper-based contact tracing efforts in Port Loko District, Sierra Leone. An Ebola Contact Tracing application called ECT app was developed to manage the communication between contact tracing coordinators and contact tracers. Our proposal focuses on tracking individuals who may have come in contact with an infected patient. We could extend our end application to enable support for coordination and cooperation.

The importance and effectiveness of epidemic contact tracing via communication traces are discussed in ~\cite{farrahi2014epidemic}. A contact tracing data set was used to show when to initiate the tracing and explains how the choice between tracing of random people from the population and the infected people influences the effectiveness of the contract tracing process. But, Farrahi~\textit{et al.}~\cite{farrahi2014epidemic} does not present any mechanisms to collect contract tracing data from the users in a privacy-preserving fashion.

Shahabi~\textit{et al.}~\cite{10.1145/2820783.2820880} presents a framework called PLACE in a vision paper to trace the proximity of individuals based on location data. ~\cite{10.1145/2820783.2820880} discussed the importance of processing the information in a privacy-preserving manner, but it did not present a specific solution. Our proposal focuses on tracing contact using obfuscated Bluetooth beacons without relying on GPS as a primary source. Besides, GPS-based approaches do not provide reliable results in an indoor environment.

EPIC~\cite{8422886} and ENACT~\cite{prasad2017enact} are developed to trace contacts using a mobile application in a privacy-preserving manner. Both these frameworks rely on a wireless access point, wherein the mobile phones are expected to report the access point identifier and the timestamp to a server, which is then matched against the infected user's information to detect contact proximity. Such an approach would only work on locations with wireless access points. Our proposal aims to trace contact with mobile phones alone.   

Qathrady~\textit{et al.}~\cite{7763193} presents an infection tracing framework for hospitals based on a centralized server. This scheme relies on static infrastructures deployed in the hospital environment, and it does not address privacy explicitly. Our proposal focuses on a location-agnostic mobile-based application.

Reddy~\textit{et al.}~\cite{DBLP:journals/corr/ReddyKRC15} presents a dengue-monitoring application, which helps the population to report dengue to the health authorities along with their location information. Besides, it includes a symptoms checker to help the users. This application relies on the information provided by the users, and it does not record communication traces. Our proposed application aims to record physical contacts using Bluetooth technology.

Yoneki~\textit{et al.}~\cite{YONEKI201483} presents EpiMap to create a map of infected people and their contacts using a mobile application called FluPhone, which collects Bluetooth proximity data and GPS location to trace connections in the case of infection. But, FluPhone does not present any privacy-preserving mechanisms. Our proposal aims to develop a mobile application that collects traces without violating the privacy of the users. 

In summary, all the existing solutions either rely on wireless access points or lack support for tracing contacts in a privacy-preserving, opt-in fashion.  

\section{CONTAIN: Privacy-oriented Contact Tracing Protocols}
\label{sec:contain}
\subsection{Design Goals}
The design goals of the CONTAIN framework are described in this section.
\begin{itemize}
    \item The application should not rely on external infrastructure such as WiFi access points with known locations or the use of GPS for determining contact information as it may violate the location privacy of the user.
    \item The application should not disclose any personally identifiable information to other mobile devices.
    \item A user should be able to check if they have been near an infected user on their own, in a completely private manner, without being forced to reveal to anyone else their infection status.  
    \item The anonymized information needed for other users to determine if they have been in contact with a user that is infected should be made available to the public through a trusted server only if the user can prove to the operator of the server (e.g. through a medical certificate) that he or she is infected.
\end{itemize}

\begin{figure}
\begin{center}
    \centerline{\includegraphics[scale=0.33]{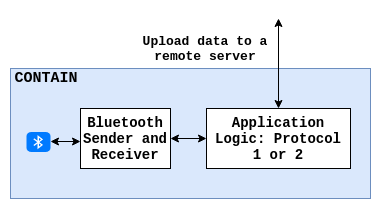}}
    \caption{Components of the CONTAIN Application.}
    \label{fig:comp}
\end{center}
\end{figure}

\subsection{System description}
Figure~\ref{fig:comp} shows the components of the CONTAIN framework.
\begin{itemize}
    \item \textbf{Mobile Device} Mobile devices (smartphones) have become prominent in the last decade. Almost every individual is carrying a highly capable mobile device (smartphone). The core processes of the CONTAIN application run on user's smartphones. 
    We describe below two different protocols that can be used for the CONTAIN application, one based on encrypted beacons (see Section~\ref{sec:proto1}) and one based on random beacons (see Section~\ref{sec:proto2}). 
    The CONTAIN application does not rely on or require GPS (in a given implementation of it, it may be possible to additionally allow users to reveal GPS information to the application on an opt-in basis, but this is not an essential part or requirement of our design). 
    \item \textbf{Bluetooth technology} Bluetooth is one of the widely used communication technology in contemporary smartphones. Today, this technology is used for short-range communication with hands-free earphones and external Bluetooth-based devices such as speakers and other audio systems. The communication range of Bluetooth is on the order of 10 meters or less~\cite{8419192}. 
    \item \textbf{Verification server} The smartphone and the Bluetooth technology can be used to gather contact traces in a private manner, which is discussed in Section~\ref{sec:proto1} and ~\ref{sec:proto2}. A third-party server is required to verify the infection status of a user (than claims to be infected) and then to make anonymous information from that infected user available for other users to use, in a private manner, to verify if they have been near the infected user.
\end{itemize}

The CONTAIN application requires Bluetooth technology in combination with computation resources and ability to access the verification server over the Internet (through any means such as Cellular or WiFi). Note that the contact tracing logic on the phone transmits and receives an anonymous encrypted or random Bluetooth beacon from other mobile devices running the CONTAIN application.

\begin{figure}
\begin{center}
    \centerline{\includegraphics[scale=0.26]{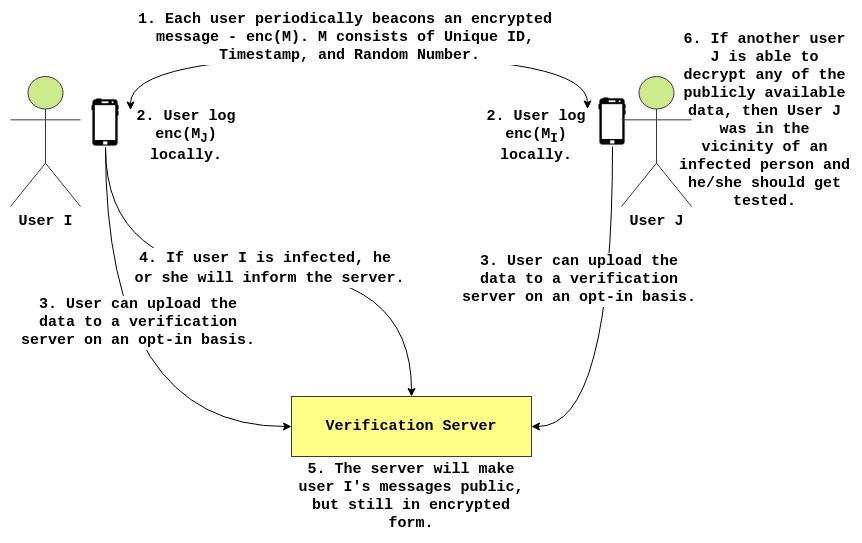}}
    \caption{Illustration of how the Privacy-Sensitive Contact Tracing App would work with our first protocol.}
    \label{fig:proto1}
\end{center}
\end{figure}


\subsection{Protocol 1: Encrypted Beacons}
\label{sec:proto1}
Figure~\ref{fig:proto1} shows the illustration of the first contact tracing protocol.
\begin{enumerate}
\item Each user periodically beacons, using Bluetooth, a message, $\mathcal{M}$, consisting of a unique name or ID, a time-stamp, and a random number (salt) that changes over time. Each message is encrypted using a symmetric key $enc(\mathcal{M})$. Note that the users do not share this key with others. 
\item Other users that hear the encrypted message beacon log it locally.
\item Either periodically or in a batch, on an opt-in basis, each user can upload all the encrypted beacon messages they have heard to a common verification server.
\item If a user $i$ becomes infected, they inform the server, on an opt-in basis, with evidence that they are infected, such as a medical report or ``infection certificate". The above step of uploading encrypted beacon messages could also be taken in conjunction with this step.  
\item The verification server proceeds to make all the messages uploaded by user $i$ publicly available, but still in the encrypted form.
\item Each other user $j$ can privately check these now publicly available encrypted messages to see if they can decrypt any of them. Note that this verification process could be automated. If any user manages to decrypt the message, then he or she is at risk as he or she has been near an infected person. In this case, the user should then proceed to get tested themselves.
\end{enumerate}

\begin{figure}
\begin{center}
    \centerline{\includegraphics[scale=0.26]{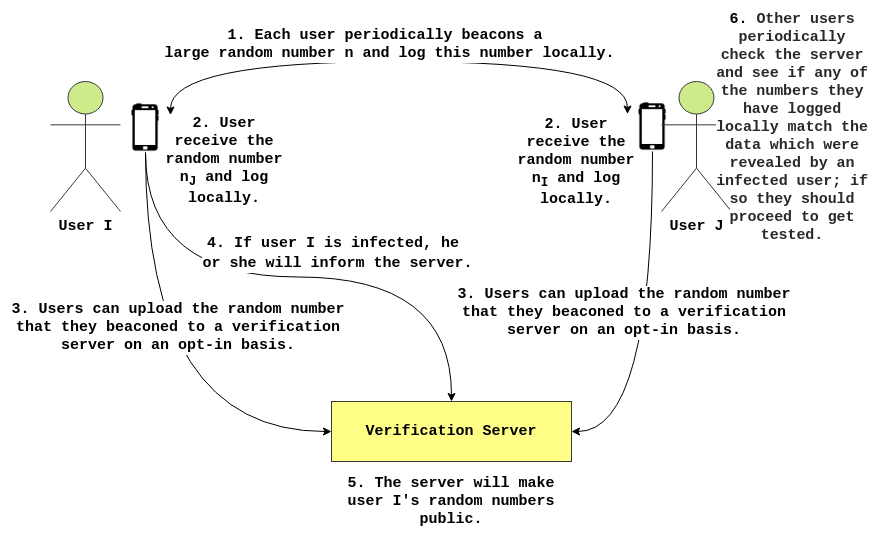}}
    \caption{Illustration of how the Privacy-Sensitive Contact Tracing App would work with our second protocol.}
    \label{fig:proto2}
\end{center}
\end{figure}

\subsection{Protocol 2: Random Beacons}
\label{sec:proto2}
Figure~\ref{fig:proto2} shows the illustration of the second contact tracing protocol.
\begin{enumerate}
    \item Each user beacons a sufficiently large random number and logs this random number locally. Here, a large number is chosen to minimize the chance of collisions with beacons generated by other users.
    \item Others that hear the beacon log this number locally.
    \item Either periodically or in a batch, on an opt-in basis, each user can upload all the random beacon messages they have transmitted to a common verification server.
    \item If a user $i$ becomes infected, on an opt-in basis, they inform the verification server with evidence that they are infected, such as a medical report or ``infection certificate”. The above step of uploading the random beacon messages could also be done in conjunction with this step. 
    \item The verification server then proceeds to make all the random numbers uploaded by this user publicly available.
    \item Other users periodically check the server and see if any of the numbers they have logged locally match those that were revealed by an infected user; if so they should proceed to get tested.
\end{enumerate}

The fourth step in both protocols requires a trusted third party that a) operates the ``publicizing” server, and b) can verify through a medical certificate that the user whose uploaded anonymous data is being made public has truly tested positive for being infected. This party must be trusted and perform the verification correctly to minimize false alarms. It may be necessary for this role to be fulfilled by an organization (whether public or private) with relevant experience and reputation. For example, it could be a healthcare organization or government entity, or medical insurance provider.  


Both protocols require the ability to beacon information over short range Bluetooth links. Today’s mobile phones, for security, privacy and energy efficiency reasons, generally require applications that do such beaconing to run in the foreground; regrettably, this could make the protocols less effective than they could be. But perhaps mobile device manufacturers could be prevailed upon to provide greater access to such an app in the public interest. Alternatively, at least at relatively more crowded venues such as airports, campuses, shopping areas or public transportation, users can be encouraged to download and turn on these apps. 

\subsection{Anonymizing an Infected User\label{sec:anonInfected}} Note that the information revealed by the verification server does not explicitly contain anything that would reveal to others who the infected user is; it only allows other users to determine if they have been near an infected user and when. In some cases, if a user knows that during a certain time they were only in contact with (within bluetooth range) of a specific person they know, they may be able to infer the identity of the infected person from the time information provided. One way the infected user could potentially avoid even this minimal chance of disclosure would be to only release information from times when they know they were in the presence of many other devices (which could be determined based on the timestamps of logged messages), and of course they could also opt-out of notifying other parts. 

While it may be assumed the infected user must make themselves known to the owner/operator of the verification server since they must provide the medical certificate that they are infected, there is an additional step that could be taken to anonymize them further. This would involve asking the infected user to use a public-key pair that does not tie them to a real-world identity, and have the medical authority digitally sign an infection certificate (using their own key pair) that only includes the infected user's public key. The infected user can then provide this digital certificate to the verification server. From this point on the verification server can verify using the public key of the medical authority that the anonymous user contacting it (whom they know only through its public key) is indeed infected, and the verification server can thus ensure that only the (anonymous) data uploaded by a user that is truly infected is being made publicly available. 

\subsection{Anonymizing Encounter Time \label{sec:anonET}}

We would like to clarify one additional aspect that is *not* guaranteed to be kept private by either protocol. If an infected user $i$ has opted-in to notify the verification server and share the data it logged, then any user $j$ that has actually been near the user $i$ will know the time of encounter, which will a) de-anonymize the location of the encounter (since the user may remember where they were at that encounter time; or \emph{a fortiori} in protocol 1 they could also inject their own GPS location explicitly into the message that is encrypted to keep track of where they were during each encounter), and b) de-anonymize the identify of the infected user $i$ to $j$ (if user $j$ knows and remembers or has some other way of identifying who they were near at the time of encounter). 

Fortunately, there is a solution to this problem. If it is desired to further strengthen the system to provide for an additional layer of privacy to avoid the above shortcoming, one possible solution would be to use protocol 2 (random beacons) and eliminate the public notification step so that the verification server does *not* publish all the random numbers sent by the infected user. Instead, require the user interested in checking if they were near an infected user to upload all the random numbers they have logged along with an anonymizing "ID" to the server (using TOR or something similar to ensure an anonymous upload). Then the server will verify if there is any overlap between the random numbers from the infected user and random numbers uploaded the user that is checking. If so, it publishes only a notification that the user corresponding to that anonymous ID has been (or not been) in contact with an infected user. 

\section{Privacy Analysis}
\label{sec:privacy}
In this section we formally identify and prove the privacy guarantees we can provide. The properties and their proofs assume that the trusted verification server behaves correctly in all cases. 

We first list the key privacy properties that both our protocols guarantee below:

\begin{itemize}
\item \textbf{P1.} Beacons emitted by a user do not reveal any personally identifying information or location information about that user to other users.
\item \textbf{P2.} Users that are not infected are not required to upload any information to the verification server.
\item \textbf{P3.} Users will not be notified by the verification server about or receive any data from the verification server that can help them verify potential contact with users that are not infected
\item \textbf{P4.} Users that are infected must opt-in in order for other users to determine if they have been near that infected user.
\item \textbf{P5.} Users can check if they have been near an (opted-in) infected user without revealing any personally identifiable information about themselves. 
\end{itemize}

Further, there are two additional privacy properties that can be shown to hold with the appropriate additional modifications as discussed in section~\ref{sec:anonInfected} (for both protocols) and section~\ref{sec:anonET} (for protocol 2 only):

\begin{itemize}
\item \textbf{P6.} Users that are infected can keep themselves anonymous even from the operator of the verification server.
\item \textbf{P7.} A user that finds out it has been contact with an infected user will not be able to determine when or where exactly the contact happened. 
\end{itemize}

 \textbf{Proof of P1:} We assume that the underlying Bluetooth protocol itself is making user of ephemeral, time-varying ID's and not sending static, identifiable MAC addresses (such schemes are already implemented by privacy-sensitive mobile device operating systems such as Apple's iOS~\cite{AStudyofMACAddressRandomizationinMobileDevicesandWhenitFails}).  In protocol 1, the beacons are encrypted using a symmetric key that is not revealed to any other user. The use of a salt that can be changed each time further makes it so that there is no way to connect a particular individual or device with the logged data since it will be different each time the salt is changed.  In protocol 2, the beacons consist of random numbers that can also change each time, so again, there is no identifying information being transmitted. Neither protocol requires collection of any location information. In protocol 1, the user could chose to encrypt their own GPS data if they so wish, but this information could only be checked by them on their own since it is encrypted with a key that only they possess. 

 \textbf{Proof of P2:} This property is a direct consequence of the two protocols. At no step is a device required to upload any information if they are not infected. Although step 3 in each protocol does allow each user to upload their random logged info periodically, they are not required to do so before they are infected, and even then it is on an opt-in basis. 

 \textbf{Proof of P3:} Since users that are not infected do not upload any information to the verification server in either protocol, there is no way for other users to be notified about or learn anything about them.

\textbf{Proof of P4:} For other users to determine if they have been near an infected user, the verification server needs to make available logs from that infected user. However, in both protocols, the logs are provided by infected users only on an opt-in basis. 

\textbf{Proof of P5:} For both protocols, the verification server can make the anonymous information they get from the infected users available on a public website. Users can use an anonymous web browser such as TOR to download that information and check if they have been near any of the infected user in a completely private manner, without revealing any of their identifying information to the operator of the verification server or any other party.

\textbf{Proof of P6:} To allow infected users to keep themselves anonymous from the verification, the certificate from the medical authority should include only the public key of the infected user, as described in the additional step of section~\ref{sec:anonInfected}. If that additional measure is taken, then the infected users gain an additional level of anonymity and privacy. 

\textbf{Proof of P7:} This is addressed by the solution presented in Section~\ref{sec:anonET}. By restricting to protocol 2 and asking users that want to check if they are infected to anonymously provide their logged data to the verification server, it can be ensured that the users that are checking will be notified that there was an encounter but not when or where the encounter happened. 


\section{Evaluation}
\label{sec:eval}
Most mobile operating systems today require apps that use Bluetooth communication to run in the foreground. Thus it requires active user intervention to be useful. In this section we evaluate through trace-based simulations how well CONTAIN would work as a function of how long the app is turned on by the user each day, and when it is turned on. We use a dataset named Asturies provided by CRAWDAD.org for our simulation~\cite{oviedo-asturies-er-20160808}. The dataset contains the encounter of mobile devices with each other at different timestamps. We split the day time into 24 slots and mapped one hour of the dataset to each time slot. Users are considered to turn on the CONTAIN mobile app to participate in the experiment according to three different scenarios, which are described in Section~\ref{Evaluation:Methodology}.

During the experiments, we varied both the number of initial infected users that are chosen randomly and also the probability that the disease gets transferred between people to see their effects on the results. Moreover,
we considered several time slots that users turn on their Bluetooth-based CONTAIN app to evaluate the effectiveness of our protocols in different scenarios. Each experiment is repeated 1000 times to capture the randomness of the simulations.


\subsection{Methodology}
\label{Evaluation:Methodology}
We conducted two sets of experiments. First we designed an experiment to study how the infection rate grows among people according to the initial number of infected users and the contagiousness probability.

Second, we designed an experiment with three different scenarios in terms of when users schedule the time slots that they turn on their Bluetooth-based CONTAIN app. 
\begin{itemize}
    \item \textbf{Random:} In this scenario, users turn on their Bluetooth randomly in 12 contiguous hours that correspond to the most active time of the day.  
    \item \textbf{Decentralized:} In this scenario, each user turns on his or her Bluetooth and runs the CONTAIN app when they are in a crowded place such as train stations, offices, shopping malls, etc. in a decentralized manner. For this scenario, we sort the time slots for each user according to the number of devices that the user connected to in that time slot. The second criterion for sorting is the crowdedness of the area that the user is. We measure the crowd level of an area by counting the number of records that we have in each time slot. Then, during an experiment where the user turns on for $k$ slots, it does so for the first $k$ slots in the sorted order.   
    \item \textbf{Centralized:} In this scenario, we assume that there is some centralized coordination mechanism that ensures that all users turn on their Bluetooth and run the CONTAIN app at specific hours of the day. In this scenario, we sort the time slots according to first the average number of active mobile devices, and second the number of records that we have in each hour in the dataset. Thus the app is activated by everyone during the hours with the most active mobile devices.  
\end{itemize}
\subsection{Results}
\begin{figure}
\begin{center}
    \centerline{\includegraphics[scale=0.32]{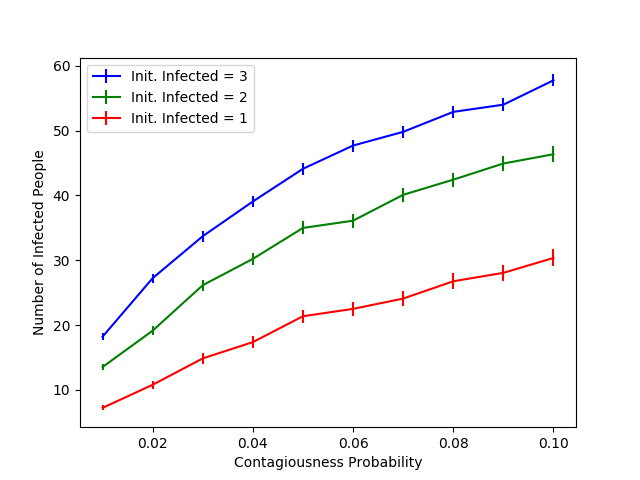}}
    \caption{Number of Infected Users vs Contagiousness Probability and Initially Infected Users}
    \label{fig:infected}
\end{center}
\end{figure}

Figure \ref{fig:infected} shows the results from our first set of experiments, showing the number of users who get infected by infectious diseases such as CoVID-19 and SARS-CoV2 with 90\% confidence interval versus contagiousness probability. These are also the number of initial users who are assumed to be infected at the beginning of the simulation. As you can see, as we have more number of initially infected users and also higher contagiousness probability, more people get infected.  

\begin{figure}
\begin{center}
    \centerline{\includegraphics[scale=0.34]{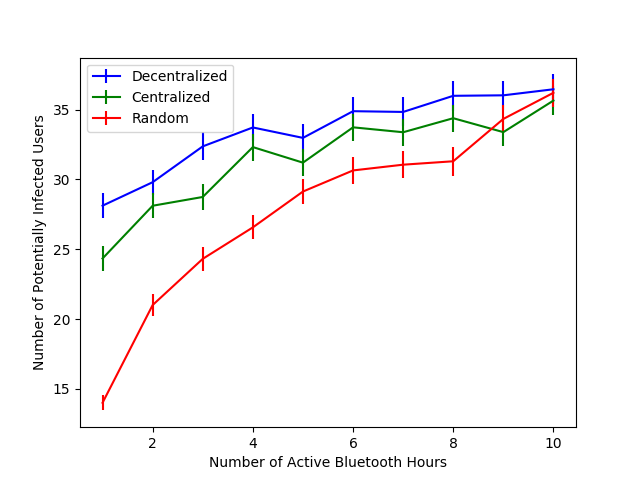}}
    \caption{Number of Test Required Users vs Number of Active Bluetooth Hours}
    \label{fig:test_req}
\end{center}
\end{figure}

Figure \ref{fig:test_req} represents the number of users who have been in contact with or in the proximity of infected persons. In this experiment, we assumed the contagiousness probability is 2\%, and the initial number of infected users is 2. From  Figure~\ref{fig:test_req}, it is clear that as more people cooperate and turn on their Bluetooth for more number of time slots, more number of potentially infected users could be identified reliably. 

As we can see in Figure \ref{fig:infected}, when the number of infected people grows so fast, it is essential to identify and notify potentially infected people to get tested. Besides, this helps the individuals who may have come in contact with the infected individual to take corrective actions, including self-isolation, to prevent him or her from spreading the infection to others.

In this experiment, we have, on average, 100 devices active per week and around 60000 records for those devices per week, which indicates a dense environment with lots of interactions among people. That is why such a small contagiousness probability and also initially infected people results in a considerable number of infected users. Consequently, as we can see in Figure \ref{fig:test_req}, our application informs many more people in case more number of them actively use the Bluetooth-based CONTAIN application. In this way, people can be alerted to get tested, take care of themselves and not spread the virus among
other people by restricting their activities and avoiding crowded areas. 


It is also interesting and important to note that when people act in a decentralized manner and turn on their Bluetooth-based app only when they go to a crowded environment, we will also have better performance as compared to the centralized and random scenarios. It is also interesting to see that decentralized scenario results in more number of people being detected as potentially infected and required to do the test as compared to the centralized scenario. This is because, if each person turns his or her Bluetooth in the places where lots of other people are available (this is evident from the dataset that we used in our study), much more number of encounters will be recorded as compared to the centralized rule, which does not fit everyone's activity.

\section{Conclusion}
\label{sec:conc}
In this paper, we have presented CONTAIN, which is a privacy-friendly and Bluetooth-based mobile application for contact tracing. Unlike the existing contact tracing application, the proposed contact tracing protocols have shown to be infrastructure independent. We have presented two contact tracing protocols, both of which do not reveal any personally identifiable information. Through a simulation study, we have shown how the contact tracing can effectively mitigate the spread of infectious diseases. 

We plan to investigate how privacy-oriented approaches influence adoption in our future work. Besides, we will also develop methods to overcome system challenges that prevent Android and iOS users from running Bluetooth-based applications in the background. And, the existing contract tracing projects do not provide any open-source software for experimentation purposes, with the exception of FluPhone~\cite{YONEKI201483}. We plan to develop and release open-source software implementations of CONTAIN for both Android and iOS devices, and perform a real world pilot study in the future. 

\bibliographystyle{ACM-Reference-Format}
\bibliography{sample-base}


\begin{thebibliography}{12}


\ifx \showCODEN    \undefined \def \showCODEN     #1{\unskip}     \fi
\ifx \showDOI      \undefined \def \showDOI       #1{#1}\fi
\ifx \showISBNx    \undefined \def \showISBNx     #1{\unskip}     \fi
\ifx \showISBNxiii \undefined \def \showISBNxiii  #1{\unskip}     \fi
\ifx \showISSN     \undefined \def \showISSN      #1{\unskip}     \fi
\ifx \showLCCN     \undefined \def \showLCCN      #1{\unskip}     \fi
\ifx \shownote     \undefined \def \shownote      #1{#1}          \fi
\ifx \showarticletitle \undefined \def \showarticletitle #1{#1}   \fi
\ifx \showURL      \undefined \def \showURL       {\relax}        \fi
\providecommand\bibfield[2]{#2}
\providecommand\bibinfo[2]{#2}
\providecommand\natexlab[1]{#1}
\providecommand\showeprint[2][]{arXiv:#2}

\bibitem[\protect\citeauthoryear{{Altuwaiyan}, {Hadian}, and
  {Liang}}{{Altuwaiyan} et~al\mbox{.}}{2018}]%
        {8422886}
\bibfield{author}{\bibinfo{person}{T. {Altuwaiyan}}, \bibinfo{person}{M.
  {Hadian}}, {and} \bibinfo{person}{X. {Liang}}.}
  \bibinfo{year}{2018}\natexlab{}.
\newblock \showarticletitle{EPIC: Efficient Privacy-Preserving Contact Tracing
  for Infection Detection}. In \bibinfo{booktitle}{\emph{2018 IEEE
  International Conference on Communications (ICC)}}. \bibinfo{pages}{1--6}.
\newblock
\showISSN{1938-1883}


\bibitem[\protect\citeauthoryear{Cabrero, García, García, and
  Melendi}{Cabrero et~al\mbox{.}}{2016}]%
        {oviedo-asturies-er-20160808}
\bibfield{author}{\bibinfo{person}{Sergio Cabrero}, \bibinfo{person}{Roberto
  García}, \bibinfo{person}{Xabiel~G. García}, {and} \bibinfo{person}{David
  Melendi}.} \bibinfo{year}{2016}\natexlab{}.
\newblock \bibinfo{title}{{CRAWDAD} dataset oviedo/asturies-er (v.
  2016-08-08)}.
\newblock \bibinfo{howpublished}{Downloaded from
  \url{https://crawdad.org/oviedo/asturies-er/20160808}}.
\newblock
\urldef\tempurl%
\url{https://doi.org/10.15783/C7302B}
\showDOI{\tempurl}


\bibitem[\protect\citeauthoryear{{Collotta}, {Pau}, {Talty}, and
  {Tonguz}}{{Collotta} et~al\mbox{.}}{2018}]%
        {8419192}
\bibfield{author}{\bibinfo{person}{M. {Collotta}}, \bibinfo{person}{G. {Pau}},
  \bibinfo{person}{T. {Talty}}, {and} \bibinfo{person}{O.~K. {Tonguz}}.}
  \bibinfo{year}{2018}\natexlab{}.
\newblock \showarticletitle{Bluetooth 5: A Concrete Step Forward toward the
  IoT}.
\newblock \bibinfo{journal}{\emph{IEEE Communications Magazine}}
  \bibinfo{volume}{56}, \bibinfo{number}{7} (\bibinfo{date}{July}
  \bibinfo{year}{2018}), \bibinfo{pages}{125--131}.
\newblock
\showISSN{1558-1896}


\bibitem[\protect\citeauthoryear{Danquah, Hasham, MacFarlane, Conteh, Momoh,
  Tedesco, Jambai, Ross, and Weiss}{Danquah et~al\mbox{.}}{2019}]%
        {danquah2019use}
\bibfield{author}{\bibinfo{person}{Lisa~O Danquah}, \bibinfo{person}{Nadia
  Hasham}, \bibinfo{person}{Matthew MacFarlane}, \bibinfo{person}{Fatu~E
  Conteh}, \bibinfo{person}{Fatoma Momoh}, \bibinfo{person}{Andrew~A Tedesco},
  \bibinfo{person}{Amara Jambai}, \bibinfo{person}{David~A Ross}, {and}
  \bibinfo{person}{Helen~A Weiss}.} \bibinfo{year}{2019}\natexlab{}.
\newblock \showarticletitle{Use of a mobile application for Ebola contact
  tracing and monitoring in northern Sierra Leone: a proof-of-concept study}.
\newblock \bibinfo{journal}{\emph{BMC infectious diseases}}
  \bibinfo{volume}{19}, \bibinfo{number}{1} (\bibinfo{year}{2019}),
  \bibinfo{pages}{810}.
\newblock


\bibitem[\protect\citeauthoryear{Farrahi, Emonet, and Cebrian}{Farrahi
  et~al\mbox{.}}{2014}]%
        {farrahi2014epidemic}
\bibfield{author}{\bibinfo{person}{Katayoun Farrahi}, \bibinfo{person}{Remi
  Emonet}, {and} \bibinfo{person}{Manuel Cebrian}.}
  \bibinfo{year}{2014}\natexlab{}.
\newblock \showarticletitle{Epidemic contact tracing via communication traces}.
\newblock \bibinfo{journal}{\emph{PloS one}} \bibinfo{volume}{9},
  \bibinfo{number}{5} (\bibinfo{year}{2014}).
\newblock


\bibitem[\protect\citeauthoryear{Martin, Mayberry, Donahue, Foppe, Brown,
  Riggins, Rye, and Brown}{Martin et~al\mbox{.}}{2017}]%
        {AStudyofMACAddressRandomizationinMobileDevicesandWhenitFails}
\bibfield{author}{\bibinfo{person}{Jeremy Martin}, \bibinfo{person}{Travis
  Mayberry}, \bibinfo{person}{Collin Donahue}, \bibinfo{person}{Lucas Foppe},
  \bibinfo{person}{Lamont Brown}, \bibinfo{person}{Chadwick Riggins},
  \bibinfo{person}{Erik~C. Rye}, {and} \bibinfo{person}{Dane Brown}.}
  \bibinfo{year}{2017}\natexlab{}.
\newblock \showarticletitle{A Study of MAC Address Randomization in Mobile
  Devices and When it Fails}.
\newblock \bibinfo{journal}{\emph{Proceedings on Privacy Enhancing
  Technologies}} \bibinfo{volume}{2017}, \bibinfo{number}{4}
  (\bibinfo{year}{2017}), \bibinfo{pages}{365 -- 383}.
\newblock


\bibitem[\protect\citeauthoryear{Organization et~al\mbox{.}}{Organization
  et~al\mbox{.}}{2014}]%
        {world2014contact}
\bibfield{author}{\bibinfo{person}{World~Health Organization} {et~al\mbox{.}}}
  \bibinfo{year}{2014}\natexlab{}.
\newblock \showarticletitle{Contact tracing during an outbreak of Ebola virus
  disease}.
\newblock  (\bibinfo{year}{2014}).
\newblock


\bibitem[\protect\citeauthoryear{Prasad and Kotz}{Prasad and Kotz}{2017}]%
        {prasad2017enact}
\bibfield{author}{\bibinfo{person}{Aarathi Prasad} {and} \bibinfo{person}{David
  Kotz}.} \bibinfo{year}{2017}\natexlab{}.
\newblock \showarticletitle{ENACT: Encounter-based Architecture for Contact
  Tracing}. In \bibinfo{booktitle}{\emph{Proceedings of the 4th International
  on Workshop on Physical Analytics}}. \bibinfo{pages}{37--42}.
\newblock


\bibitem[\protect\citeauthoryear{{Qathrady}, {Helmy}, and
  {Almuzaini}}{{Qathrady} et~al\mbox{.}}{2016}]%
        {7763193}
\bibfield{author}{\bibinfo{person}{M.~A. {Qathrady}}, \bibinfo{person}{A.
  {Helmy}}, {and} \bibinfo{person}{K. {Almuzaini}}.}
  \bibinfo{year}{2016}\natexlab{}.
\newblock \showarticletitle{Infection tracing in smart hospitals}. In
  \bibinfo{booktitle}{\emph{2016 IEEE 12th International Conference on Wireless
  and Mobile Computing, Networking and Communications (WiMob)}}.
  \bibinfo{pages}{1--8}.
\newblock
\showISSN{null}


\bibitem[\protect\citeauthoryear{Reddy, Kumar, Rollings, and Chandra}{Reddy
  et~al\mbox{.}}{2015}]%
        {DBLP:journals/corr/ReddyKRC15}
\bibfield{author}{\bibinfo{person}{Emmenual Reddy}, \bibinfo{person}{Sarnil
  Kumar}, \bibinfo{person}{Nicholas Rollings}, {and} \bibinfo{person}{Rohitash
  Chandra}.} \bibinfo{year}{2015}\natexlab{}.
\newblock \showarticletitle{Mobile Application for Dengue Fever Monitoring and
  Tracking via {GPS:} Case Study for Fiji}.
\newblock \bibinfo{journal}{\emph{CoRR}}  \bibinfo{volume}{abs/1503.00814}
  (\bibinfo{year}{2015}).
\newblock
\showeprint[arxiv]{1503.00814}


\bibitem[\protect\citeauthoryear{Shahabi, Fan, Nocera, Xiong, and Li}{Shahabi
  et~al\mbox{.}}{2015}]%
        {10.1145/2820783.2820880}
\bibfield{author}{\bibinfo{person}{Cyrus Shahabi}, \bibinfo{person}{Liyue Fan},
  \bibinfo{person}{Luciano Nocera}, \bibinfo{person}{Li Xiong}, {and}
  \bibinfo{person}{Ming Li}.} \bibinfo{year}{2015}\natexlab{}.
\newblock \showarticletitle{Privacy-Preserving Inference of Social
  Relationships from Location Data: A Vision Paper}. In
  \bibinfo{booktitle}{\emph{Proceedings of the 23rd SIGSPATIAL International
  Conference on Advances in Geographic Information Systems}} (Seattle,
  Washington) \emph{(\bibinfo{series}{SIGSPATIAL ’15})}.
  \bibinfo{publisher}{Association for Computing Machinery},
  \bibinfo{address}{New York, NY, USA}, Article \bibinfo{articleno}{9},
  \bibinfo{numpages}{4}~pages.
\newblock
\showISBNx{9781450339674}


\bibitem[\protect\citeauthoryear{Yoneki and Crowcroft}{Yoneki and
  Crowcroft}{2014}]%
        {YONEKI201483}
\bibfield{author}{\bibinfo{person}{Eiko Yoneki} {and} \bibinfo{person}{Jon
  Crowcroft}.} \bibinfo{year}{2014}\natexlab{}.
\newblock \showarticletitle{EpiMap: Towards quantifying contact networks for
  understanding epidemiology in developing countries}.
\newblock \bibinfo{journal}{\emph{Ad Hoc Networks}}  \bibinfo{volume}{13}
  (\bibinfo{year}{2014}), \bibinfo{pages}{83 -- 93}.
\newblock
\showISSN{1570-8705}
\newblock
\shownote{(1)Special Issue : Wireless Technologies for Humanitarian Relief \&
  (2)Special Issue: Models And Algorithms For Wireless Mesh Networks.}


\end{thebibliography}

\end{document}